\documentstyle[prd,aps,preprint]{revtex}
\tightenlines
\newcommand {\be}{\begin{eqnarray}}
\newcommand {\ee}{\end{eqnarray}}
\input epsf.tex
\def\DESepsf(#1 width #2){\epsfxsize=#2 \epsfbox{#1}}
\begin{document}

\draft
%@twocolumnfalse\endcsname
\preprint{\vbox{
\hbox{UMD-PP-99-109}
}}
\title{ Sterile Neutrinos in $E_6$ and a Natural Understanding of 
Vacuum Oscillation Solution to the Solar Neutrino Puzzle}
\author{ Z. Chacko and R. N. Mohapatra }

\address{Department of
Physics, University of Maryland, College Park, MD, 20742}
\date{May, 1999}
\maketitle
\begin{abstract}
If Nature has chosen the vacuum oscillation solution to the Solar neutrino
puzzle, a key theoretical challenge is to understand the extreme smallness
of the $\Delta m^2_{\nu_e-\nu_X}$ ($\sim 10^{-10}$ eV$^2$) required for
the purpose. We find that
in a class of models such as $[SU(3)]^3$ or its parent group $E_6$, which
contain one sterile neutrino, $\nu_{is}$ for each family, the $\Delta
m^2_{\nu_i-\nu_{is}}$ is proportional to the cube of the lepton Yukawa
coupling. Therefore fitting the atmospheric neutrino data then predicts
the $\nu_e-\nu_{es}$ mass difference square to be
$\sim \left(\frac{m_e}{m_{\mu}}\right)^3 \Delta m^2_{atmos}$, where the
atmospheric neutrino data is assumed to be solved via the
$\nu_{\mu}-\nu_{\mu s}$ oscillation. This provides a natural explanation
of the vacuum oscillation solution to the solar neutrino problem. 
 \end{abstract}

\section{Introduction}
As is widely known by now, the Super-Kamiokande data has provided
conclusive evidence for the
existence of oscillations of the muon neutrinos from cosmic rays
\cite{superK}. While it is yet to be determined what final state the 
cosmic ray $\nu_{\mu}$s oscillate to ($\nu_{\tau}$ or $\nu_{\mu s}$), it
is known
that the mixing angle is near maximal and the $\Delta m^2_{atmos}\sim
10^{-3}$ eV$^2$. Similarly the solar neutrino data from Super-Kamiokande
and other experiments\cite{solar} are also making a very convincing case
for oscillations of the electron neutrinos emitted by the Sun in order to 
understand the observed deficit of the solar neutrinos\cite{bahcall}.
Again, in this case also, it is not clear what final state the $\nu_e$
oscillates to on its way from the Sun to Earth. It could either be
$\nu_{\mu}$ or $\nu_{es}$. There are however several mixing angle and
mass difference possibilities in the solar case\cite{bahcall}. One of the
possibilities is the vacuum oscillation of $\nu_e-\nu_\mu$ or
$\nu_e-\nu_{es}$ type. In order to explain the observations one needs in
this case that the $\Delta m^2_{\nu_e-\nu_X}\sim 10^{-10}$ eV$^2$ and a
maximal mixing like in the atmospheric neutrino case. The recent
indications of a seasonal dependence of the solar neutrino events in the
708 day Super-Kamiokande data\cite{modu} would seem
to support this explanation although it is by no means the
only way to understand it\cite{concha}. If the vacuum oscillation 
explanation finally wins, then a serious theoretical challenge is to
understand the unusually small mass difference squared between the
neutrinos needed for the purpose. It is the goal of this letter to
propose a way to answer this challenge within a gauge theory framework.

The first observation that motivates our final scenario is the symmetry
between the solution to the atmospheric neutrino data and the vacuum
oscillation solution to the solar neutrino data in that the mixing angles
are maximal. This might suggest a generation independence of the neutrino
mixings patterns. An implementation of such an idea would naturally
require that in each case i.e. solar as well as atmospheric the active
neutrinos (i.e. $\nu_e$ and $\nu_{\mu}$) oscillate into the sterile
neutrinos\cite{sterile} to be denoted by $\nu_{es}$ and $\nu_{\mu s}$
respectively. The complete three family picture would then require that
there be one sterile neutrino per family. One class of models that
lead to such a scenario\cite{lew} is the mirror universe\cite{bere}
picture where  
the particles and the forces in the standard model are duplicated in a 
mirror symmetric manner. There is no simple way to understand the ultra
small $\Delta m^2$ needed for the vacuum oscillation solution in this
case. In this letter we focus on an alternative
scheme based on the grand unification group $[SU(3)]^3$ or its parent
group $E_6$. 

We find it convenient to use the $E_6$ notation. As is
well-known\cite{gursey}, under the SO(10) group, the {\bf 27}-dimensional
representation of $E_6$ decomposes to ${\bf 16}_{+1}\oplus {\bf 10}_{-2}
\oplus {\bf 1}_{+4}$ where the subscripts represent the U(1) charges. The
{\bf 16} is well known to contain the left and the right handed neutrinos
(to be denoted by us as $\nu_i$ and $\nu^c_i$, $i$ being the family
index). The {\bf 10} contains two neutral colorless fermions which behave
like neutrinos but are $SU(2)_L$ doublets and the last neutral colorless
fermion in the {\bf 27}, which we identify as the sterile neutrino is the
one contained in {\bf 1} (denoted by $\nu_{is}$). In general in this
model, we will have for each generation a $5\times 5$ ``neutrino'' mass
matrix and we will show how the small masses for the sterile neutrino and
the known neutrino come out as a consequence of a generalized seesaw
mechanism. Furthermore, we will see how as a consequence of the smallness
of the Yukawa couplings of the standard model, we will not only get
maximal mixing between the active and the sterile neutrinos of each
generation but also the necessary ultra-small $\Delta m^2$ needed in the
vacuum oscillation solution without fine tuning of
parameters\footnote{An $E_6$ model for the neutrino puzzles was
first discussed 
by Ma\cite{ma}, where his goal was to understand the smallness of the
sterile neutrino masses. Our model is different in many respects and
addresses the question of maximal mixing, small $\Delta m^2$'s as well 
as the small neutrino masses. Our picture also differs from other recently
proposed models\cite{roy}}. The way this comes
about in our model is that to the lowest order in the Yukawa couplings,
the $\nu_i$ and $\nu_{is}$ form a Dirac neutrino with a mass
proportional to the generational Yukawa coupling
$\lambda_i$ of the corresponding generation. However they become
pseudo-Dirac to order $\lambda^3_i$ leading to nearly degenerate
neutrinos with a mass
splitting $\Delta m^2_i\approx \lambda^3_i$. Therefore fixing the 
$\Delta m^2_{atmos}$ gives the right value for the $\Delta m^2_e$ needed
for the vacuum oscillation solution.

Let us now present the basic idea of the model for one generation of
neutrinos consisting of $\nu_i, \nu_{is}$. Suppose that their mass matrix
is given by the following $2\times 2$ matrix:
\begin{eqnarray}
M_i= m_{0i}\left(\begin{array}{cc}
\lambda^2_i & \lambda_i \bar{f}_i \\
\lambda_i \bar{f}_i & \lambda^2_i \bar{\epsilon}_i\end{array}\right)
\end{eqnarray}
Since $\lambda_i \ll 1$, it is clear that the two neutrinos are maximally
mixed with a mass $m_i\simeq \lambda_i \bar{f}_i m_{0i}$ amd 
$\Delta m^2_{i}\simeq
\lambda^3_i \bar{f}_i m^2_{0i}$ provided $\bar{\epsilon} \le 1$ 
and $\bar{f} \ge 1$. These relations are true generation by generation.
We shall show in the next section that a mass matrix of this form emerges
naturally from $E_6$ and its subgroup $[SU(3)]^3$ with $\bar{f}_i\approx
1$. The main 
difference between these two groups arises from the fact that for the
simplest models based on $E_6$ the Yukawa coupling $\lambda_i$ is
necessarily related to the 
Yukawa coupling of the corresponding up type quark. This is not true 
for $[SU(3)]^3$, for which $\lambda_i$ is expected to be related to the 
Yukawa coupling of the corresponding lepton.   

Now let us look at the atmospheric neutrino data. Since $\Delta
m^2_{2}\simeq 2- 5\times 10^{-3}$ eV$^2$, if we choose the $\nu_{\mu}$
mass to be of order $0.2-0.5$ eV (anticipating that we want to accomodate 
the LSND data\cite{lsnd}), then we get $\lambda_2\simeq 10^{-2}$, 
which is a typical second generation Yukawa coupling. Note that this is a
plausible value even for $[SU(3)]^3$ since
in supersymmetric models $m_{\mu}\simeq \lambda_2 v_d$ and $v_d$ can be
considerably less ( e.g 10 GeV for a $tan \beta \simeq 24$) than the
standard model value of 246 GeV for the
symmetry breaking parameter. Since the same formula applies to the
$\nu_e$ and $\nu_{es}$ sector, assuming no large flavour dependence in 
the coefficients $f_i$ and $\epsilon_i$, all we need
in order to predict their masses and mass
differences is the value for $\lambda_1/\lambda_2$. Irrespective of 
whether we consider $E_6$ or $[SU(3)]^3$
 we find $\lambda_1/\lambda_2 \simeq
5\times 10^{-3}$. This leads to a value for the $\Delta m^2_1 \simeq
2\times 10^{-10}$ eV$^2$, which is clearly of the right order of
magnitude. Our main point is not to insist on precise numbers but rather
to illustrate the idea that a cubic dependence of the neutrino mass
difference squares on the generational Yukawa couplings to leptons of the
standard model can lead to an understanding of the extreme smallness
of $\Delta m^2$ value needed in the vacuum oscillation solution to
the solar neutrino puzzle\cite{ma2}.

Extending our idea to the third generation, we find that value of the
$\nu_{\tau}$ mass is $(\lambda_3/\lambda_2) 0.2$ eV. This imples 
$m_{\nu_{\tau}}\approx 2-3$ eV for $[SU(3)]^3$ 
 which is interesting for cosmology since
this would mean that about 10-15\% of the mass of universe could come from
neutrinos. This expectation can eventually be tested when the finer
measurements of the angular power spectrum is carried out in the MAP and
PLANCK experiments in the next few years. However for minimal versions of  
$E_6$, $\lambda_3/\lambda_2 \sim m_t/m_c$ yielding a
value $m_{\nu_{\tau}}\approx 20-30$ eV which is unacceptable for a 
realistic cosmology. This means that the simplest $E_6$ model that can
accomodate our scenario is one where the quark lepton symmetry is
broken. The other possibility is to have some flavour dependence on the
coefficients 
$\bar{f}_i$ and $m_{0i}$. The extent of required flavour dependence
is certainly not extreme and we consider models based on
both groups as 
realistic candidates for a complete theory of neutrino masses.  

\section{The Model}

Let us now proceed to construct the mass matrix in Eq. (1) in the context
of an $E_6$ model.
As usual, we will assign matter to the {\bf 27} dimensional representation 
of the
group and we have already noted that there are five neutrino-like fields
in the model which will mix among each other subsequent to symmetry
breaking. It is therefore necessary to describe the symmetry breaking of
$E_6$. To implement the symmetry breaking we use three pairs of 
${\bf 27}+\bar{\bf27}$ representations and one {\bf 78}-dim. field.
The pattern of symmetry breaking is as follows:

1) $<27_1>$ and $<\overline{27}_1>$ have GUT scale vevs in the SO(10)
   singlet direction.
   
2) $<27_{16}>$ and $<\overline{27}_{16}>$ have GUT scale vevs in the 
   $\nu$ and $\nu^c$ directions respectively. They break SO(10) down 
   to SU(5).
 
3) The $<78_{[1,45]}>$ completes the breaking of SU(5) to the standard model
   gauge group at the GUT scale. We assume the VEVs reside both in the 
   adjoint and in the singlet of SO(10). 
   
4) $<27_{10}>$ and $<\overline{27}_{10}>$ contain the Higgs doublets of the 
   MSSM. It is assumed that $H_u$ and $H_d$ are both linear combinations
   arising partially from the $<27_{10}>$ and partially from the 
   $<\overline{27}_{10}>$

In addition to the above there is another field labelled by $27^\prime$
whose $\nu^c$ component mixes with a singlet S and one linear combination 
of this pair (denoted by $S^\prime$)
remains light below the GUT scale. As a consequence of 
radiative symmetry breaking  this picks up a VEV at the electroweak
scale. We will show later how this can occur. The remaining components 
of $27^\prime$ have GUT scale mass. 

Let us now write down the relevant terms in the superpotential that lead
to a $5\times 5$ ``neutrino'' mass matrix of the form we desire.
 To keep matters simple let us
ignore generation mixings, which can be incorporated very trivially.

\begin{eqnarray}
W~=~\lambda_i \psi_i\psi_i 27_{10}  + f_i \psi_i\psi_i 27^\prime + 
\frac{\alpha_i}{M_{P\ell}}\psi_i\psi_i 27_{1} 78_{[1,45]}  
+\frac{\gamma_i}{M_{P\ell}}\psi_i\psi_i \overline{27}_{16} \overline{27}_{16}
\end{eqnarray}
We have chosen only a subset of allowed terms in the theory and believe
that it is reasonable to assume a discrete symmetry (perhaps in the
context of a string model) that would allow only this subset.
In any case since we are dealing with a supersymmetric theory, radiative
corrections will not generate any new terms in the superpotential.

Note that in Eq. (2), since it is the first term that leads to lepton and
quark masses of
various generations, it carries a generation label and obeys a
hierarchical pattern, whereas the $f_i$'s not being connected to known
fermion masses need not obey a hierarchical pattern. We will from now on
assume that each $f_i\approx 1$, and see where it leads us.

After substituting the VEVs for the Higgs fields in the above equation,
we find a $5\times 5$ mass matrix\footnote{Although the form of
this mass matrix is same as in \cite{ma}, the results of our paper are
different.} of the following form for the
neutral lepton fields of each generation in the basis $(\nu, \nu_s, \nu^c,
E^0_u, E^0_d)$:
\begin{eqnarray}  
M~=~ \left(\begin{array}{ccccc}
0 & 0 & \lambda_i v_{u} & f_i v' & 0 \\
0 & 0 & 0 & \lambda_i v_{d} & \lambda_i v_{u} \\
\lambda_i v_{u} & 0 & M_{\nu^c,i} & 0 & 0 \\
f_i v' & \lambda_i v_{d} & 0 & 0 & M_{10,i} \\
0 & \lambda_i v_{u} & 0 & M_{10,i} & 0 \end{array} \right)
\end{eqnarray}
Here $M_{\nu^c,i}$ is the mass of the right handed neutrino and $M_{10,i}$
is the mass of the entire {\bf 10}-plet in the {\bf 27} matter multiplet.
Since {\bf 10} contains two full SU(5) multiplets, gauge coupling unification
will not be effected even though its mass is below the GUT scale.

Note that the $ 3\times 3$ mass matrix involving the $(\nu^c, E^0_u,
E^0_d)$ have superheavy entries and will therefore decouple at low
energies. Their effects on the spectrum of the light neutrinos will be
dictated by the seesaw mechanism\cite{grsyms}. The light neutrino mass
matrix involving $\nu_i, \nu_{is}$ can be written down as:
\begin{eqnarray}
M_{light}\simeq \frac{1}{M_{\nu^c,i}} \left(\begin{array}{ccc}
 \lambda_i v_{u} & f_i v' & 0 \\
 0 & \lambda_i v_{d} & \lambda_i v_{u}\end{array}\right)
\left(\begin{array}{ccc}
 1 & 0 & 0 \\
 0 & 0 & \epsilon \\
 0 & \epsilon & 0 \end{array} \right)\left(\begin{array}{cc}
\lambda_i v_{u} & 0 \\
f_i v' & \lambda_i v_{d}  \\
0 & \lambda_i v_{u}  \end{array} \right)
\end{eqnarray}

where $\epsilon_i = M_{10,i}/ M_{\nu^c,i}$. Note that $\epsilon_i $ is
expected to
be of order one. This leads to the $2\times 2$ mass matrix for the $(\nu,
\nu_c)$ fields of each generation which is of the form in Eq. (1),

\begin{eqnarray}
M_i= m_{0i}\left(\begin{array}{cc}
\lambda^2_i & \lambda_i \bar{f}_i\\
\lambda_i \bar{f}_i & \lambda^2_i \bar{\epsilon_i}\end{array}\right)
\end{eqnarray}

Here $ m_{0i} = \frac{v_{u}^2}{M_{\nu^c,i}}$, $\bar{f}_i = f_i\epsilon_i
v'/v_u$,
and $ \bar{\epsilon_i} = 2 \epsilon_i cot{\beta}$. Taking $M_{Pl} \sim
10^{19}GeV$, $M_{GUT} \sim 10^{16}$ and reasonable values of the unknown
parameters e.g. $\alpha_i\approx 
0.1$, $\gamma_i\approx 0.1$,
$f_i\approx 1$, $v'\approx v_u$, we get
$m_{0i}\simeq 20$ eV and $\epsilon\approx 1$ which leads us
to the desired pattern of masses and mass differences outlined in the
introduction. 

A crucial assumption in our analysis is that that one of the Higgs fields
has a vev along $\nu^c$ direction with a low scale (the $v'$ above). We
will now demonstrate what kind of a superpotential can lead to such a
situation.

Consider
\begin{eqnarray}
W = M 27^{\prime}\overline{27}^{\prime} + S \overline{27}^{\prime} 27_{16}
\end{eqnarray}

Since $ 27_{16} $ has a VEV, this implies that one linear combination
of
S and the $\nu^c$ component of $27^{\prime}$ (denoted by $S^\prime$)
remains light while everything 
else in  $27^{\prime}$ and $\overline{27}^{\prime}$
become heavy. If in addition the superpotential 
contains the couplings

\begin{eqnarray}
W = S 27_{10} \overline{27}_{10} + S^3
\end{eqnarray}

since $27_{10}$ and $\overline{27}_{10}$ have electroweak scale VEVs the 
light combination of S and $27^{\prime}$ ($S^\prime$) also picks up an
electroweak scale 
VEV from the trilinear soft supersymmetry breaking terms. Note that this 
is inevitable as long as electroweak symmetry is broken because such a 
trilinear term then becomes a linear term in the potential for $S^\prime$ 
and hence $S^\prime$ must pick up a VEV. We thus see that it is possible
to
get vev for the singlet field $\nu^c$ in the desired {\bf 27}-plet of the
order of the electroweak scale.

Let us next address the question of the generation mixing. We will assume
that it parallels that in the quark sector i.e. the mixing angles to start
with are small. Since the neutrino mixings have an additional contribution
coming from their seesaw mechanism, we can easily have them be smaller
than
the corresponding quark mixings. This is for instance what one would like
in order to fit the LSND data. We do not get into the details of this
since clearly it does not effect the main point of the paper.

Let us end with a few comments on the phenomenological and cosmological
implications of the model. The most severe test of this model will come
from the understanding of big bang nucleosynthesis\cite{sarkar}. Our model
within the standard assumptions that go into the discussion of BBN would
imply $N_{\nu}=6$ i.e. three extra neutrinos. However, in models with
sterile neutrinos, possibilities of large lepton asymmetry at the BBN era 
has been discussed\cite{fv}.

The second point that needs emphasizing is that in our model, both the
solar and atmospheric neutrinos involve separate sterile neutrinos in the
final state. There are well known tests\cite{vissani} of such models for
the atmospheric neutrino oscillations\cite{yasuda} where one looks for
neutral pion production. For solar neutrinos, our model is testable by the 
neutral current measurement planned for the SNO experiment\cite{sno}.

In conclusion, in this paper we have pointed out a simple way to
understand theoretically challenging possibility of a tiny mass difference
squared that may arise if the solar neutrino puzzle is to be solved via
the vacuum oscillation solution. We exploit an apparent symmetry
between the solar and the atmospheric case arising from the maximality of
mixing angles to suggest that the ultra small $\Delta m^2_{solar}$ may be
undestandable
in models of $E_6$ type, which automatically contain a sterile neutrino
in each {\bf 27} that also contains other known particles of each
generation and where the generational neutrino mass difference squared may
be proportional to the cube of the lepton Yukawa couplings. In this model 
we can also accomodate the indication for neutrino oscillations from LSND.

 This work has been supported by the National Science Foundation
grant under no. PHY-9802551 .

\end{document}